\documentclass[aps,showpacs,twocolumn]{revtex4}
\usepackage{epsfig}
\usepackage{graphicx,color,dcolumn}
\usepackage{epstopdf}
\usepackage{amsmath,amssymb}
\usepackage{multirow}
\usepackage{diagbox}
\usepackage{changes}
\usepackage{booktabs}
\usepackage{threeparttable}

\begin{document}

\title{$QQ\bar{q}\bar{q}$ in a chiral constituent quark model}

\author{Yue Tan}\email{181001003@stu.njnu.edu.cn}
\author{Weichang Lu}\email{161002001@stu.njnu.edu.cn}
\author{Jialun Ping}\email{jlping@njnu.edu.cn(Corresponding author)}

\affiliation{Department of Physics and Jiangsu Key Laboratory for Numerical
Simulation of Large Scale Complex Systems, Nanjing Normal University, Nanjing 210023, P. R. China}

\begin{abstract}
Inspired by $\Xi_{cc}$ reported by LHCb Collaboration and $X(5568)$ reported by $D0$ Collaboration,
the $QQ\bar{q}\bar{q}$ ($Q=c,b$, $q=u,d$) tetraquark states, are studied in the present work. With the
help of gaussian expansion method, two structures, diquark-antiquark and meson-meson, with all possible
color configurations are investigated systematically in a chiral quark model to search for the possible stable
states. The results show that there is no bound state in the isovector $QQ\bar{q}\bar{q}$ system, while
there are rather deep bound states in the isoscalar $bb\bar{q}\bar{q}$, $cc\bar{q}\bar{q}$ and $bc\bar{q}\bar{q}$ systems.
Mixing two structures of diquark-antidiquark and meson-meson can introduce more attractions and convert some unbound
isoscalar states into shallow bound states. The large mass of the heavy quark is beneficial to the formation of the bound
state. The separations between quarks are calculated to unravel the spacial structure of the system.
\end{abstract}

%\pacs{~}

\maketitle

\section{Introduction} \label{introduction}
Since the exotic state $X(3872)$ first observed by Belle collaboration \cite{1Choi:2003ue}, Belle and other collaborations have
reported a lot of ``$XYZ$" particles \cite{2Aaij:2019evc,3Ablikim:2013mio,4Ablikim:2013wzq}, which stimulated many researches on
hadron spectrum. It's well known that quantum chromodynamics (QCD) is the underlying approach of strong interaction, and
in principle QCD allows the existence of exotic states: multiquark states, hybrid states and glueball. However, due to
the nonperturbtive properties of QCD in the low energy region, it is unavailable for us to use it to study the hadron structures
and the hadron-hadron interactions directly. The study of exotic states can provide much essential information, which is absent
in the ordinary $qqq$ baryons and $q\bar{q}$ mesons, on low energy QCD.

In 2016, D0 collaboration observed a narrow structure in the $B_{s}^{0}\pi^{\pm}$ invariant mass spectrum with $5.1\sigma$
significance. Its mass and width are $M=5567.8\pm2.9_{-1.9}^{+0.9}$ MeV and $\Gamma=21.9\pm6.4_{-2.5}^{+5}$ MeV \cite{5D0:2016mwd}.
Because of the $B_{s}^{0}\pi^{\pm}$ decay mode, $X(5568)$ was interpreted as $su\bar{b}\bar{d}$ or $sd\bar{b}\bar{u}$ tetraquark
state. But LHCb collaboration got negative result about $X(5568)$ \cite{6LHCb:2016ppf}. Nevertheless, the D0 collaboration¡¯s new
result still insisted on the existence of this tetraquark $X(5568)$ \cite{14Abazov:2017poh}. Inspired by the discussions on $X(5568)$,
its partner state with quark content $bs\bar{u}\bar{d}$ is proposed afterwards \cite{10Chen:2018hts,11Huang:2019otd}.
One year later, LHCb collaboration reported a double charmed baryon $\Xi_{cc}^{++}$ with masss $M=3621.4\pm0.78$ MeV, which 100 MeV
heavier than what SELEX collaboration reported \cite{8Mattson:2002vu}. Which is close to our group's results \cite{9Yang:2008zzi,Yang:2017qan}.
The existence of $\Xi_{cc}^{++}$ may imply the existence of the stable $QQ\bar{q}\bar{q}$ system.

Although lacking experimental information regarding to $QQ\bar{q}\bar{q}$ system, theoretical researches on this topic have a long
history \cite{ManoharNPB,ZPC57,Moinester,Gelman,Janc,QiangZhao,Carlson,Vijande,10Chen:2018hts,11Huang:2019otd,12Karliner:2017qjm,13Eichten:2017ffp,Yang:2017qan}. Manohar and Wise obtained a weakly bound two-meson state $bb\bar{q}\bar{q}$ using one-pion exchange \cite{ManoharNPB}.
Moinester proposed to search for the doubly charmed tetraquarks experimentally as early as the 1990s~\cite{Moinester}.
Yang {\em et al.} systematically studied $QQ\bar{q}\bar{q}$ system by using several versions of quark models, and pointed out that
iso-scalar $bb\bar{q}\bar{q}$, $cc\bar{q}\bar{q}$ states are deep bound states, and $ss\bar{q}\bar{q}$ states are scattering
states \cite{Yang:2017qan}. Karliner {\em et al.} estimated the energy of $QQ$, $Qq$ and $qq$
from the experimental data of $Q\bar{Q}$, $Q\bar{q}$ and $q\bar{q}$ mesons according to relation between quark-quark and quark-antiquark.
Based on this method, they got $\Xi_{cc}^{++}$'s mass which was very close to experimental value and predicted the existence of $QQ\bar{q}\bar{q}$ \cite{12Karliner:2017qjm}. Eichten {\em et al.} likened lowest-lying tetraquark configuration as atom ($QQ$ is the "nuclus" while $\bar{q}\bar{q}$ is electron). By heavy-quark symmetry, they predicted that $bb\bar{q}\bar{q}$ must be stable system \cite{13Eichten:2017ffp}.
Recently Caram\'{e}s {\em et al.} got two bound states in the $bc\bar{q}\bar{q}$ system by two different methods \cite{Caramees:2018oue}.
In addition, after replacing $b\bar{q}q\bar{s}$ by $qq\bar{s}\bar{b}$, Chen {\em et al.} got a bound state in the chiral
quark model~\cite{10Chen:2018hts}. Huang {\em et al.} used  quark delocalization color screening model which intermediate-range
attraction is provided by quark delocalization and color screening to study $qq\bar{s}\bar{b}$ and $qs\bar{q}\bar{b}$ system,
and they found that the tetraquarks composed of $qq\bar{s}\bar{b}$ is more possible to form bound states than the one composed of $qs\bar{q}\bar{b}$~\cite{11Huang:2019otd}. Very recently, Yang studied $QQ\bar{q}\bar{q}$ system in a chiral quark model
by using complex scaling method, several bound states and resonance states were obtained~\cite{GYang2020}.

The QCD inspired quark models successfully describe hadron spectrum, of which the chiral quark model is most popular.
In this paper, we use chiral constituent quark model to systematically calculate $QQ\bar{q}\bar{q}$ with the help of
gaussian expansion method. Different from the other's work \cite{12Karliner:2017qjm,13Eichten:2017ffp}, we consider two kinds of
structures, meson-meson and diquark-antiquark, and their mixing. All the possible color and spin configurations are also taken
into account. The necessity for mixing the two different structures is that it is not economic way to use one structure to form a
complete set of states because all the possible excited states have to be included. Coupling the important structures to enlarge
the model space is a good choice for few-quark systems in the low-energy region.
In addition, the root mean square distance between quarks/antiquarks are calculated to unravel the structure of the states if they
are bound ones.

The paper is organized as follows. In section II, the chiral quark model and the wave-function of $QQ\bar{q}\bar{q}$
systems are presented. The numerical results are given in Sec. III. The last section is devoted to the summary of the present work.

\section{Chiral quark model and wave-function of $QQ\bar{q}\bar{q}$ system} \label{wavefunction and chiral quark model}
\subsection{Chiral quark model}
The chiral quark model has been successful both in describing the hadron spectra and hadron-hadron interactions.
The details of the model can be found in Ref. \cite{10Chen:2018hts,17Yang:2009zzp,18Vijande:2004he,19Chen:2016npt}.
The Hamiltonian of the chiral quark model consists of quarks mass, kinetic energy, and three kinds of potentials,
color confinement, one-gluon-exchange and Goldstone boson exchange.
The Hamiltonian for four-quark system is written as,
\begin{eqnarray}
H &=& \sum_{i=1}^4
m_i+\frac{p_{12}^2}{2\mu_{12}}+\frac{p_{34}^2}{2\mu_{34}}
  +\frac{p_{1234}^2}{2\mu_{1234}}  \nonumber \\
  & & +\sum_{i<j=1}^4 \left( V_{ij}^{G}+V_{ij}^{C}+\sum_{\chi=\pi,K,\eta, \sigma} V_{ij}^{\chi} \right), \nonumber\\
\end{eqnarray}
Where $m$ is the constituent masse of quark(antiquark), and $\mu$ is the reduced masse of two interacting quarks or
quark-clusters.
\begin{eqnarray}
\mu_{ij}&=&\frac{m_{{i}}  m_{{j}}}{m_{{i}} + m_{{j}}},   \nonumber\\
\mu_{1234}&=&\frac{(m_1+m_2)(m_3+m_4)}{m_1+m_2+m_3+m_4},
\end{eqnarray}
and $p_{ij}=\frac{m_jp_i-m_ip_j}{m_i+m_j}$, $p_{1234}=\frac{(m_3+m_4)p_{12}-(m_1+m_2)p_{34}}{m_1+m_2+m_3+m_4}$.

The first potential is the color confinement, the quadratic form is used here,
\begin{equation}
V_{ij}^{C}= ( -a_{c} r_{ij}^{2}-\Delta) \boldsymbol{\lambda}_i^c \cdot \boldsymbol{\lambda}_j^c .
\end{equation}
The second potential is the effective smeared one-gluon exchange interaction,
\begin{eqnarray}
 V_{ij}^{G}&=& \frac{\alpha_s}{4} \boldsymbol{\lambda}_i^c \cdot \boldsymbol{\lambda}_{j}^c
\left[\frac{1}{r_{ij}}-\frac{2\pi}{3m_im_j}\boldsymbol{\sigma}_i\cdot
\boldsymbol{\sigma}_j
  \delta(\boldsymbol{r}_{ij})\right]   \\
 & &  \delta{(\boldsymbol{r}_{ij})}=\frac{e^{-r_{ij}/r_0(\mu_{ij})}}{4\pi r_{ij}r_0^2(\mu_{ij})},~~ r_{0}(\mu_{ij})=\frac{r_0}{\mu_{ij}}.
  \nonumber
\end{eqnarray}
The third potential is the Goldstone boson exchange, coming from the effects of the chiral symmetry spontaneous breaking
of QCD in low-energy region.
\begin{eqnarray}
V_{ij}^{\pi}&=& \frac{g_{ch}^2}{4\pi}\frac{m_{\pi}^2}{12m_im_j}
  \frac{\Lambda_{\pi}^2}{\Lambda_{\pi}^2-m_{\pi}^2}m_\pi v_{ij}^{\pi}
  \sum_{a=1}^3 \lambda_i^a \lambda_j^a,  \nonumber \\
V_{ij}^{K}&=& \frac{g_{ch}^2}{4\pi}\frac{m_{K}^2}{12m_im_j}
  \frac{\Lambda_K^2}{\Lambda_K^2-m_{K}^2}m_K v_{ij}^{K}
  \sum_{a=4}^7 \lambda_i^a \lambda_j^a,  \nonumber \\
V_{ij}^{\eta} & = &
\frac{g_{ch}^2}{4\pi}\frac{m_{\eta}^2}{12m_im_j}
\frac{\Lambda_{\eta}^2}{\Lambda_{\eta}^2-m_{\eta}^2}m_{\eta}
v_{ij}^{\eta} \nonumber \\
 && \left[\lambda_i^8 \lambda_j^8 \cos\theta_P
 - \lambda_i^0 \lambda_j^0 \sin \theta_P \right],  \nonumber \\
V_{ij}^{\sigma}&=& -\frac{g_{ch}^2}{4\pi}
\frac{\Lambda_{\sigma}^2}{\Lambda_{\sigma}^2-m_{\sigma}^2}m_\sigma
\left[
 Y(m_\sigma r_{ij})-\frac{\Lambda_{\sigma}}{m_\sigma}Y(\Lambda_{\sigma} r_{ij})\right] \nonumber \\
 v_{ij}^{\chi} & = & \left[ Y(m_\chi r_{ij})-
\frac{\Lambda_{\chi}^3}{m_{\chi}^3}Y(\Lambda_{\chi} r_{ij})
\right]
\boldsymbol{\sigma}_i \cdot\boldsymbol{\sigma}_j, \nonumber \\
& & Y(x)  =   e^{-x}/x .
\end{eqnarray}
In the above formula, $\boldsymbol{\sigma}$ are the $SU(2)$ Pauli matrices; $\boldsymbol{\lambda}$, $\boldsymbol{\lambda}^{c}$
are $SU(3)$ flavor, color Gell-Mann matrices, respectively; $\alpha_{s}$ is an effective scale-dependent running coupling,
\begin{equation}
 \alpha_s(\mu_{ij})=\frac{\alpha_0}{\ln\left[(\mu_{ij}^2+\mu_0^2)/\Lambda_0^2\right]},
\end{equation}
All the parameters are determined by fitting the meson spectrum, from light to heavy, taking into account only a quark-antiquark component.
They are shown in Table~\ref{modelparameters}.

\begin{table}[t]
\begin{center}
\caption{Quark Model Parameters ($m_{\pi}=0.7$ fm, $m_{\sigma}=3.42$ fm, $m_{\eta}=2.77$ fm, $m_{K}=2.51$ fm).\label{modelparameters}}
\begin{tabular}{cccc}
\hline\noalign{\smallskip}
Quark masses   &$m_u=m_d$(MeV)     &313  \\
               &$m_{s}$(MeV)         &536  \\
               &$m_{c}$(MeV)         &1728 \\
               &$m_{b}$(MeV)         &5112 \\
\hline
Goldstone bosons   &$\Lambda_{\pi}=\Lambda_{\sigma}(fm^{-1})$     &4.2  \\
                   &$\Lambda_{\eta}=\Lambda_{K}(fm^{-1})$     &5.2  \\
                   &$g_{ch}^2/(4\pi)$                &0.54  \\
                   &$\theta_p(^\circ)$                &-15 \\
\hline
Confinement        &$a_{c}$(MeV)     &101 \\
                   &$\Delta$(MeV)       &-78.3 \\
                   &$\mu_{c}$(MeV)       &0.7 \\
\hline
OGE                 & $\alpha_{0}$        &3.67 \\
                   &$\Lambda_{0}(fm^{-1})$ &0.033 \\
                  &$\mu_0$(MeV)    &36.976 \\
                   &$\hat{r}_0$(MeV)    &28.17 \\
\hline
\end{tabular}
\end{center}
\end{table}

\subsection{The wave-function of $QQ\bar{q}\bar{q}$ system}

The $QQ\bar{q}\bar{q}$ system has two structures, meson-meson and diquark-antidiquark, and the wave function of each structure
all consists of four parts: orbit, spin, flavor and color wave functions. In addition, the wave function of each part is constructed
by coupling two sub-clusters wave functions. Thus, the wave function for each channel will be the tensor product of orbit
($|R_{i}\rangle$), spin ($|S_{j}\rangle$), color ($|C_{k}\rangle$) and flavor ($|F_{l}\rangle$) components,
\begin{equation}\label{bohanshu}
|ijkl\rangle={\cal A} |R_{i}\rangle\otimes|S_{j}\rangle\otimes|C_{k}\rangle\otimes|F_{l}\rangle
\end{equation}
${\cal A}$ is the antisymmetrization operator.

\subsubsection{orbit wave function}
The total wave function consists of two sub-clusters orbit wave functions and the relative motion wave function between two sub-clusters.
\begin{eqnarray}\label{spatialwavefunctions}
|R_{1}\rangle & = & \left[[\Psi_{l_1}({\bf r}_{12})\Psi_{l_2}({\bf
r}_{34})]_{l_{12}}\Psi_{L_r}({\bf r}_{1234}) \right]_{LM_{L}} \nonumber\\
|R_{2}\rangle&=&\left[[\Psi_{l_1}({\bf r}_{13})\Psi_{l_2}({\bf
r}_{24})]_{l_{12}}\Psi_{L_r}({\bf r}_{1324}) \right]_{LM_{L}}.
\end{eqnarray}
Where the bracket "[~]" indicates angular momentum coupling, and the "L" means total orbit angular momentum coupled by $L_r$,
relative motion angular momentum, and "$l_{12}$" coupled by "$l_1$" and "$l_2$", sub-cluster angular momenta.
In addition, we use "$|R_{1}\rangle$" denotes meson-meson structure while "$|R_{2}\rangle$" denotes diquark-antidiquark structure.
In GEM, the radial part of spatial wave function is expanded by Gaussians~\cite{20Hiyama:2003cu}:
\begin{subequations}
\label{radialpart}
\begin{align}
R(\mathbf{r}) & = \sum_{n=1}^{n_{\rm max}} c_{n}\psi^G_{nlm}(\mathbf{r}),\\
\psi^G_{nlm}(\mathbf{r}) & = N_{nl}r^{l}
e^{-\nu_{n}r^2}Y_{lm}(\hat{\mathbf{r}}),
\end{align}
\end{subequations}
where $N_{nl}$ are normalization constants,
\begin{align}
N_{nl}=\left[\frac{2^{l+2}(2\nu_{n})^{l+\frac{3}{2}}}{\sqrt{\pi}(2l+1)}
\right]^\frac{1}{2}.
\end{align}
$c_n$ are the variational parameters, which are determined dynamically. The Gaussian size parameters are chosen
according to the following geometric progression
\begin{equation}\label{gaussiansize}
\nu_{n}=\frac{1}{r^2_n}, \quad r_n=r_1a^{n-1}, \quad
a=\left(\frac{r_{n_{\rm max}}}{r_1}\right)^{\frac{1}{n_{\rm
max}-1}}.
\end{equation}
This procedure enables optimization of the ranges using just a small number of Gaussians.

\subsubsection{spin wave function}
Because of no difference between spin of quark and antiquark, the meson-meson structure has the same total spin as
the diquark-antidiquark structure. The spin wave functions of the cluster are shown below.
\begin{eqnarray}
 & & \chi_{11}^{\sigma}=\alpha\alpha,~~
\chi_{10}^{\sigma}=\frac{1}{\sqrt{2}}(\alpha\beta+\beta\alpha),~~
\chi_{1-1}^{\sigma}=\beta\beta, \nonumber \\
 & & \chi_{00}^{\sigma}=\frac{1}{\sqrt{2}}(\alpha\beta-\beta\alpha),
\end{eqnarray}
According to Clebsch-Gordan coefficient table, total spin wave function can be written below.
\begin{eqnarray}
|S_{1}\rangle & = & \chi_{0}^{\sigma1}=\chi_{00}^{\sigma}\chi_{00}^{\sigma}, \nonumber \\
|S_{2}\rangle & = & \chi_{0}^{\sigma2}= \sqrt{\frac{1}{3}}(\chi_{11}^{\sigma}\chi_{1-1}^{\sigma}
  -\chi_{10}^{\sigma}\chi_{10}^{\sigma}+\chi_{1-1}^{\sigma}\chi_{11}^{\sigma}), \nonumber \\
|S_{3}\rangle & = & \chi_{1}^{\sigma1}=\chi_{00}^{\sigma}\chi_{11}^{\sigma}, \\
|S_{4}\rangle & = & \chi_{1}^{\sigma2}=\chi_{11}^{\sigma}\chi_{00}^{\sigma}, \nonumber \\
|S_{5}\rangle & = & \chi_{1}^{\sigma3}=\frac{1}{\sqrt{2}}(\chi_{11}^{\sigma}\chi_{10}^{\sigma}-\chi_{10}^{\sigma}\chi_{11}^{\sigma}),\nonumber \\
|S_{6}\rangle & = & \chi_{2}^{\sigma1}=\chi_{11}^{\sigma}\chi_{11}^{\sigma}. \nonumber
\end{eqnarray}
Where the subscript of "$\chi_{S}^{\sigma i}$" denotes total spin of the tretraquark, and the superscript is the index of
the spin function with fixed $S$.

\subsubsection{flavor wave function}
The flavor wave functions of the sub-clusters for two structures are shown below,
\begin{eqnarray}
&& \chi_{\frac{1}{2}\frac{1}{2}}^{fm}=Q\bar{d},~~\chi_{\frac{1}{2}-\frac{1}{2}}^{fm}=-Q\bar{u},~~~~Q=b,c,s \\
&& \chi_{00}^{fd1}=\frac{1}{\sqrt{2}}\left( \bar{u}\bar{d}-\bar{d}\bar{u} \right),~~\chi_{00}^{fd2}=QQ,~~~~Q=b,c,s \nonumber \\
&& \chi_{11}^{fd}=\bar{d}\bar{d},~~\chi_{10}^{fd}=-\frac{1}{\sqrt{2}}\left( \bar{u}\bar{d}+\bar{d}\bar{u} \right),
~~\chi_{1-1}^{fd}=\bar{u}\bar{u}.
\end{eqnarray}
Where the subscripts of $\chi_{II_z}^{fm(d)i}$ are the isospin and its third component, and superscripts denote the structure and the
index (if needed). The total flavor wave functions can be written as,
\begin{eqnarray}
|F_{1}\rangle & = & \chi_{0}^{fm1}=\frac{1}{\sqrt{2}}\left( \chi_{\frac{1}{2}\frac{1}{2}}^{fm}
 \chi_{\frac{1}{2}-\frac{1}{2}}^{fm}-\chi_{\frac{1}{2}-\frac{1}{2}}^{fm} \chi_{\frac{1}{2}\frac{1}{2}}^{fm} \right), \nonumber\\
|F_{2}\rangle & = & \chi_{1}^{fm2}=\chi_{\frac{1}{2}\frac{1}{2}}^{fm}  \chi_{\frac{1}{2}\frac{1}{2}}^{fm}, \\
|F_{3}\rangle & = & \chi_{0}^{fd1}=\chi_{00}^{fd2}\chi_{00}^{fd1}, \nonumber \\
|F_{4}\rangle & = & \chi_{1}^{fd2}=\chi_{00}^{fd2}\chi_{11}^{fd}. \nonumber
\end{eqnarray}
Where the subscript of $\chi_{I}^{fm(d)i}$ is total isospin.

\subsubsection{color wave function}
The colorless tetraquark system has four color structures, including $1\otimes1$, $8\otimes8$, $3\otimes \bar{3}$ and $6\otimes \bar{6}$,
\begin{eqnarray}
|C_{1}\rangle & = & \chi_{1\otimes1}^{m1}=\frac{1}{\sqrt{9}}(\bar{r}r\bar{r}r+\bar{r}r\bar{g}g+\bar{r}r\bar{b}b
   +\bar{g}g\bar{r}r+\bar{g}g\bar{g}g \nonumber \\
  & + & \bar{g}g\bar{b}b+\bar{b}b\bar{r}r+\bar{b}b\bar{g}g+\bar{b}b\bar{b}b), \nonumber \\
|C_{2}\rangle & = & \chi_{8\otimes8}^{m2}=\frac{\sqrt{2}}{12}(3\bar{b}r\bar{r}b+3\bar{g}r\bar{r}g+3\bar{b}g\bar{g}b
   +3\bar{g}b\bar{b}g \nonumber \\
 &+ & 3\bar{r}g\bar{g}r+ 3\bar{r}b\bar{b}r+2\bar{r}r\bar{r}r+2\bar{g}g\bar{g}g+2\bar{b}b\bar{b}b-\bar{r}r\bar{g}g \nonumber \\
&-& \bar{g}g\bar{r}r-\bar{b}b\bar{g}g-\bar{b}b\bar{r}r-\bar{g}g\bar{b}b-\bar{r}r\bar{b}b). \\
|C_{3}\rangle & = & \chi^{d1}_{\bar{3}\otimes 3} =\frac{\sqrt{3}}{6}(rg\bar{r}\bar{g}-rg\bar{g}\bar{r}+gr\bar{g}\bar{r}
    -gr\bar{r}\bar{g}+rb\bar{r}\bar{b}, \nonumber \\
&- & rb\bar{b}\bar{r}+br\bar{b}\bar{r}-br\bar{r}\bar{b}+gb\bar{g}\bar{b}-gb\bar{b}\bar{g}+bg\bar{b}\bar{g}-bg\bar{g}\bar{b}), \nonumber \\
|C_{4}\rangle & = & \chi^{d2}_{6\otimes \bar{6}}=\frac{\sqrt{6}}{12}(2rr\bar{r}\bar{r}+2gg\bar{g}\bar{g}+2bb\bar{b}\bar{b}
    +rg\bar{r}\bar{g} \nonumber \\
&+ &rg\bar{g}\bar{r}+gr\bar{g}\bar{r}+gr\bar{r}\bar{g}+rb\bar{r}\bar{b}+rb\bar{b}\bar{r}+br\bar{b}\bar{r} \nonumber \\
&+ &br\bar{r}\bar{b}+gb\bar{g}\bar{b}+gb\bar{b}\bar{g}+bg\bar{b}\bar{g}+bg\bar{g}\bar{b}).\nonumber
\end{eqnarray}
To write down the wave functions easily for each structure, the different orders of the particles are used. However, when coupling the
different structure, the same order of the particles should be used.

\subsubsection{total wave function}
In the present work, $QQ\bar{q}\bar{q}$ and $QQ^{\prime}\bar{q}\bar{q}~(Q^{\prime}\ne Q)$ systems are all investigated.
The antisymmetrization operators are different for different systems. For $QQ\bar{q}\bar{q}$ system, we have
\begin{equation}
{\cal A}=1-(13)-(24)+(13)(24)
\end{equation}
for meson-meson structure, and
\begin{equation}
{\cal A}=1-(12)-(34)+(12)(34)
\end{equation}
for diquark-antidiquark structure. For $QQ^{\prime}\bar{q}\bar{q}$ system, the antisymmetrization operator becomes
\begin{equation}
{\cal A}=1-(24)
\end{equation}
for meson-meson structure, and
\begin{equation}
{\cal A}=1-(34)
\end{equation}
for diquark-antidiquark structure. After applying the antisymmetrization operator, some wave function will vanish, which means
that the states are forbidden. For example, $IJ^P=00^{+}$ $bb\bar{q}\bar{q}$ state is a forbidden state. All of allowed channels
are listed in Table \ref{channel}.
\begin{table}[ht]
\centering
\fontsize{7}{7}\selectfont
\caption{All of allowed channels}.\label{channel}
\begin{tabular}{cccccccc}
\hline \hline
\multicolumn{4}{c}{$QQ\bar{q}\bar{q}$}& \multicolumn{4}{c}{$QQ^{\prime}\bar{q}\bar{q}$}  \\ \hline
  $IJ^{P}$ & channel  & $IJ^{P}$ & channel         & $IJ^{P}$ & channel          & $IJ^{P}$ &  channel  \\ \hline
  $00^{+}$ & ~~~~~~~  & $10^{+}$ & $|1112\rangle $ & $00^{+}$ & $ |1111\rangle $ & $10^{+}$ & $ |1112\rangle $\\
           & ~~~~~~~  & ~~~~~~   & $|1122\rangle $ & ~~~~~~~~ & $ |1121\rangle $ & ~~~~~~~~ & $ |1122\rangle $\\
   ~~~~~~  & ~~~~~~~  & ~~~~~~   & $|1212\rangle $ & ~~~~~~~~ & $ |1211\rangle $ & ~~~~~~~~ & $ |1212\rangle $\\
   ~~~~~~  & ~~~~~~~  & ~~~~~~   & $|1222\rangle $ & ~~~~~~~~ & $ |1221\rangle $ & ~~~~~~~~ & $ |1222\rangle $\\
   ~~~~~~  & ~~~~~~~  & ~~~~~~   & $|2144\rangle $ & ~~~~~~~~ & $ |2133\rangle $ & ~~~~~~~~ & $ |2144\rangle $\\
   ~~~~~~  & ~~~~~~~  & ~~~~~~   & $|2234\rangle $ & ~~~~~~~~ & $ |2243\rangle $ & ~~~~~~~~ & $ |2234\rangle $\\
   \hline
  $01^{+}$ & $|1311\rangle $ & $11^{+}$ & $|1312\rangle $ & $01^{+}$ & $ |1311\rangle $ & $11^{+}$ & $ |1312\rangle $\\
  ~~~~~~~~ & $|1321\rangle $ & ~~~~~~~~ & $|1322\rangle $ & ~~~~~~~~ & $ |1321\rangle $ & ~~~~~~~~ & $ |1322\rangle $\\
  ~~~~~~~~ & $|1411\rangle $ & ~~~~~~~~ & $|1412\rangle $ & ~~~~~~~~ & $ |1411\rangle $ & ~~~~~~~~ & $ |1412\rangle $\\
  ~~~~~~~~ & $|1421\rangle $ & ~~~~~~~~ & $|1422\rangle $ & ~~~~~~~~ & $ |1421\rangle $ & ~~~~~~~~ & $ |1422\rangle $\\
  ~~~~~~~~ & $|1511\rangle $ & ~~~~~~~~ & $|1512\rangle $ & ~~~~~~~~ & $ |1511\rangle $ & ~~~~~~~~ & $ |1512\rangle $\\
  ~~~~~~~~ & $|1521\rangle $ & ~~~~~~~~ & $|1522\rangle $ & ~~~~~~~~ & $ |1521\rangle $  & ~~~~~~~~& $ |1522\rangle $\\
  ~~~~~~~~ & $|2343\rangle $ & ~~~~~~~~ & $|2534\rangle $ & ~~~~~~~~ & $ |2343\rangle $ & ~~~~~~~~ & $ |2334\rangle $\\
  ~~~~~~~~ & $|2433\rangle $ & ~~~~~~~~ & ~~~~~~~~~~~~~~~ & ~~~~~~~~ & $ |2433\rangle $ & ~~~~~~~~ & $ |2534\rangle $\\
  ~~~~~~~~ & ~~~~~~~~        & ~~~~~~~~ & ~~~~~~~~        & ~~~~~~~~ & $ |2543\rangle $ & ~~~~~~~~ & $ |2444\rangle $\\
  \hline
  $02^{+}$ & ~~~~~~~~        & $12^{+}$ & $|1612\rangle $ & $02^{+}$ & $ |1611\rangle $ & $12^{+}$ & $ |1612\rangle $\\
  ~~~~~~~~ & ~~~~~~~~        & ~~~~~~~~ & $|1622\rangle $ & ~~~~~~~~ & $ |1621\rangle $ & ~~~~~~~~ & $ |1622\rangle $\\
  ~~~~~~~~ & ~~~~~~~~        & ~~~~~~~~ & $|2634\rangle $ & ~~~~~~~~ & $ |2643\rangle $ & ~~~~~~~~ & $ |2634\rangle $\\
  \hline\hline
\end{tabular}
\end{table}

\section{Results}

In the present calculation, we are interested in the possible bound states. all the angular momenta are set to 0.
The single channel and channel coupling calculations are performed. For the bound state, the root mean square distances
between quark-quark/antiquark are also given to unravel the structure of the state.

\subsection{$QQ\bar{q}\bar{q}$ system}
In the iso-scalar section, the states with quantum number $IJ^P=00^{+}$ and $02^{+}$ are forbidden states, only $01^{+}$ states
are allowed. The results for iso-scalar $QQ\bar{q}\bar{q}$ states with $IJ^P=01^{+}$ are shown in Tables \ref{QQ1} and \ref{distance}.
In the table \ref{QQ1}, the second column gives the energies in the single channel calculation. The third column gives the percentages of
each channel for the lowest eigen-state which energy is given in the row marked by ``c.c." in the channel coupling calculation.
The last row gives the thresholds for $QQ\bar{q}\bar{q}~(Q=b,c,s)$ systems.

\begin{figure}[htp]
%   \begin{center}
   \epsfxsize=9.5cm\epsffile{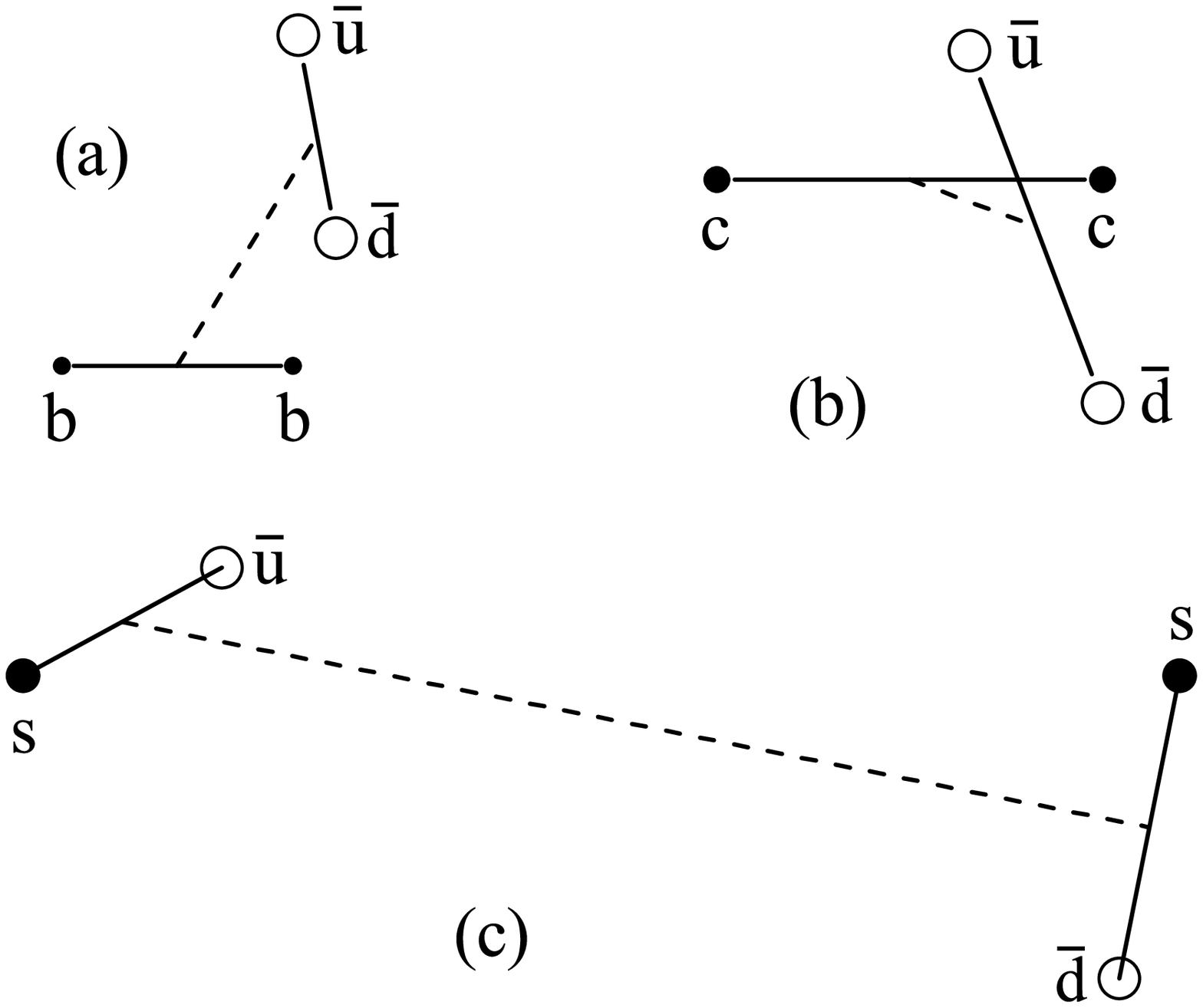}
    \caption{the spacial structures of $QQ\bar{q}\bar{q}$ system with $IJ^P=01^{+}$.}\label{struct}
%   \end{center}

  \centering
%   % Requires \usepackage{graphicx}
   \includegraphics[width=9cm]{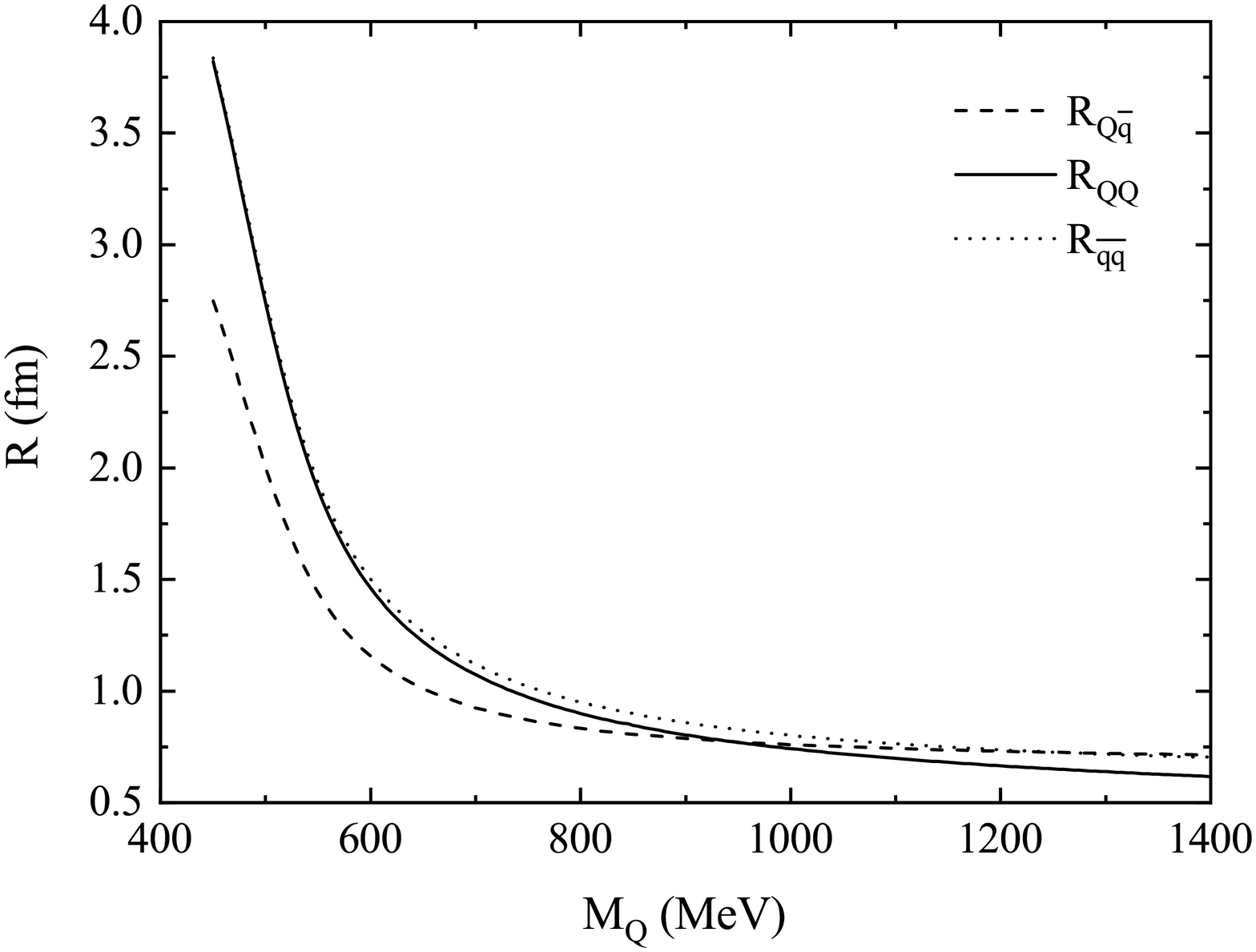}\\
   \caption{The variation of the distances between quarks (antiquarks) with the mass of heavy quark.}\label{masschange}
\end{figure}

For $bb\bar{q}\bar{q}$ states,
there are three channels have energies lower than the threshold in the single-channel calculation, the lowest one appears
in the diquark-antidiquark structure with the light antidiquark being ``good antidiquark" \cite{Jaffe}. The channel-coupling pushes down
the lowest energy 20 MeV and pushes other states above the the threshold. The binding energy of the bound state is $\sim 311$ MeV.
The percentages of channels in the lowest eigen-state show that the main component is diquark-antidiquark structure with color configuration
$\bar{3}\otimes 3$. The distances between quarks (antiquarks) (see Table \ref{distance} tell us again that the bound state has
diquark-antidiquark structure, $r_{QQ}=0.3$ fm, $r_{\bar{q}\bar{q}}=0.4$ fm and $r_{Q\bar{q}}=0.6$ fm.
For $cc\bar{q}\bar{q}$ states,
there is one channel has energy lower than the threshold in the single-channel calculation, it is the diquark-antidiquark structure
with color configuration $\bar{3}\otimes 3$. The channel-coupling shifts the lowest energy 30 MeV downwards. We obtain the binding energy
of the bound state $\sim 182$ MeV. The percentages of channels in the lowest eigen-state and the distances between particles advocate
that the dominant structure of the bound state is the diquark-antidiquark structure with color configuration $\bar{3}\otimes 3$.
For $ss\bar{q}\bar{q}$ states,
all the channels have energies larger than the threshold in the single-channel calculation. However, the channel-coupling plays an
important role and it leads to emergence of a bound state, with binding energy $\sim 9$ MeV. The percentages of channels in
the lowest eigen-state and the distances between quarks (antiquarks) claim that the bound state is a molecule,
$r_{QQ}=2.10$ fm, $r_{\bar{q}\bar{q}}=2.13$ fm and $r_{Q\bar{q}}=1.58$ fm. One can see that the distance between $s$ and $q$
is larger in four-quark system than that in the meson $Q\bar{q}$. The reason for the large separation is due to the antisymmetrization,
If we give up the antisymmetrization, it is reasonable here because the separation between two $Q$'s ($\bar{q}$'s) is large,
the distance between $Q$ and $\bar{q}$ in one cluster is 0.53 fm.
To illustrate the results for $QQ\bar{q}\bar{q}~(Q=b,c,s)$ systems with $IJ^P=01^+$, the spacial structures of the systems
are shown in Fig.~\ref{struct}. From the figure, we can see that the systems take the 3-dimensional structure, and prefers the
diquark-antidiquark structure with $Q=b,c$ (Fig.1(a),1(b)), and the size of the four-quark object increases with the decreasing
mass of $Q$. When the distance between quarks (antiquarks) is larger than the confinement scale ($\sim$ 1 fm), the system is
transferred to meson-meson structure (Fig. 1(c)). To show what is the critical heavy quark mass for this ``phase transition",
the variation of separations between particles with the heavy quark mass $m_Q$ is presented in Fig.~\ref{masschange}. The critical
heavy quark mass can be read from Fig.~\ref{masschange}, $m_Q (\mbox{critical})\sim 600$ MeV. It is a second-order phase transition.

\begin{table}[tp]
\centering
%\fontsize{8}{8}\selectfont
\caption{The energies of iso-scalar $QQ\bar{q}\bar{q}~(Q=s,c,b)$ system with $IJ^P=01^+$. ``c.c." stands for channel-coupling.}\label{QQ1}
\begin{tabular}{ccccccc}
\hline \hline
channel & \multicolumn{2}{c}{$bb\bar{q}\bar{q}$}&\multicolumn{2}{c}{$cc\bar{q}\bar{q}$}&\multicolumn{2}{c}{$ss\bar{q}\bar{q}$}\\ \hline
    $|1311\rangle$&10590.2&$ 8.07\%$&3843.8&$ 9.18\%$&1410.9&$ 42.92\%$\\
    $|1321\rangle$&10765.6&$ 0.87\%$&4168.6&$ 0.96\%$&2060.0&$ 0.08\%$\\
    $|1411\rangle$&10590.2&$ 8.07\%$&3843.8&$ 9.18\%$&1410.9&$ 42.92\%$\\
    $|1421\rangle$&10765.6&$ 0.87\%$&4168.6&$ 0.96\%$&2060.0&$ 0.08\%$\\
    $|1511\rangle$&10629.6&$ 24.70\%$&3961.7&$ 9.58\%$&1830.2&$ 3.94\%$\\
    $|1521\rangle$&10738.3&$ 3.31\%$&4102.2&$ 3.37\%$&1903.3&$ 1.03\%$\\
    $|2343\rangle$&10763.2&$ 0.01\%$&4134.6&$ 0.05\%$&2048.5&$ 0.57\%$\\
    $|2433\rangle$&10304.6&$ 54.09\%$&3709.3&$66.71\%$&1636.4&$ 8.46\%$\\
c.c.           &\multicolumn{2}{c}{10282.9}&\multicolumn{2}{c}{3660.8}&\multicolumn{2}{c}{1398.6}\\
Threshold      &\multicolumn{2}{c}{10600.5}&\multicolumn{2}{c}{3843.1}&\multicolumn{2}{c}{1407.7}\\\hline
\end{tabular}
\end{table}

\begin{table}[tp]
\centering
\caption{The distances between particles for the lowest eigen-states of $QQ\bar{q}\bar{q}~(Q=s,c,b)$ system with $IJ^P=01^+$ (unit: fm).}\label{distance}
\begin{tabular}{ccccccc}
\hline\hline
      & $r_{12}$    & $r_{13}$    & $r_{14}$    & $r_{23}$    & $r_{24}$    & $r_{34}$ \\ \hline
~~$b\bar{q}b\bar{q}$~~ & ~~0.6~~   & ~~0.3~~ & ~~0.6~~ & ~~0.6~~ & ~~0.4~~ & ~~0.6~~  \\
$c\bar{q}c\bar{q}$ & 0.7   & 0.5 & 0.7 & 0.7 & 0.6 & 0.7  \\
$s\bar{q}s\bar{q}$ & 1.58   & 2.10 & 1.58 & 1.58 & 2.13 & 1.58  \\
\hline\hline
\end{tabular}
\end{table}

To find which interaction leads to the form of bound states, the contribution of each terms in the system hamiltonian to
the total energy of the system are tabulated in Table \ref{masscon}. The one-gluon-exchange interaction is separated into
two terms, color Coulomb (Coul) and color magnetic interaction (CMI). $V^{K}$ makes no contribution to the system which is
not listed in the table. From the table, we can see that the major contribution to attraction of the system is from
$\pi$-meson exchange potential, the major contribution to repulsion is the kinetic energy.
Because the compactness increases with the increase of heavy quark mass, the Goldstone-boson-exchange will become strong
when the mass of heavy quark goes high. However, the strong attraction from the Goldstone-boson-exchange will be neutralized
by the kinetic energies, which will be more repulsive when the $m_Q$ becomes large. Fortunately the color-magnetic interaction
always contributes the sizable attraction for the heavier system, and for the lighter system, other terms, $V^{\eta}$ and
$V^{\sigma}$ supplements the attraction to bind the system.

In the iso-vector section, all the states with quantum number $IJ^P=00^{+}$, $01^+$ and $02^{+}$ are allowed states. The results
of channel coupling calculation are shown in Tables \ref{QQ2}. The energies of all the eigen-states are above the thresholds,
no bound state is found. Comparing with iso-scalar states, the light antidiquark in iso-vector states is ``bad antidiquark",
the color-magnetic interaction in the one-gluon-exchange is repulsive in this case, which destroy the antidiquark structure,
so un-binding of states is understandable.

\begin{table}[tp]
 \centering
 \fontsize{8}{8}\selectfont
 \caption{Contributions of each terms in Hamiltonian to the energy of the system. ``c.c." donates the full channel coupling and ``t.m."
  means the two free mesons. (unit:MeV).}\label{masscon}
  \begin{tabular}{cccccccccc}
  \hline \hline
    $m_Q$  &          & T      & $V^C$   & $V^{\text{Coul}}$   & $V^{\text{CMI}}$  & $V^{\pi}$   & $V^{\eta}$  & $V^{\sigma}$ \\\hline
    1400   & c.c.     & 1457.7 &$-419.6$ & $-648.9$ & $-361.2$ & $-437.1$ &  64.1 & $-39.0$ \\
           & t.m.     & 951.0  &$-392.1$ & $-631.4$ & $-165.2$ &  0.0  &   0.0&  0.0\\
           &$\Delta_{E}$&$+506.7$& $-27.5$& $-17.5$ & $-196.0$ & $-437.1$ & $+64.1$ & $-39.0$\\\hline
     800   & c.c.     &1492.8 & $-389.3$& $-675.2$ & $-446.6$& $-292.1$ &  24.2 & $-24.1$ \\
           & t.m.     &1203.8& $-385.3$ & $-680.6$ & $-375.1$ &   0.0&  0.0&  0.0\\
           &$\Delta_{E}$&$+289.0$& $-4.0$ & $ +5.4$& $-71.5$ & $-292.1$ &  $+24.2$ & $-24.1$ \\\hline
     500   & c.c.     &1583.8 & $-365.6$ & $-724.5$ & $-654.9$ & $-71.5$ &  $-50.8$ & $-6.0$ \\
           & t.m.     &1522.9& $-366.3$ & $-728.4$ & $-652.1$&   0.0& $-67.5$ & 0.0\\
           &$\Delta_{E}$& $+50.9$ &  $-0.7$ &  $+3.9$ &  -2.8& -71.5&  $-16.7$ & $-6.0$ \\\hline
  \end{tabular}
\end{table}
\begin{table}[tp]
  \centering
  \caption{The energies of iso-vector $QQ\bar{q}\bar{q}~(Q=s,c,b)$ systems (unit: MeV).}\label{QQ2}
    \begin{tabular}{ccccc}
    \hline \hline
     state& $IJ^{P}$ & c.c. &  Threshold & ~~~status~~~ \\ \hline
     $bb\bar{q}\bar{q}$&$10^{+}$&~~10561.9~~&~~10561.4~~&~~ub \\
                       &$11^{+}$&~~10600.8~~&~~10600.5~~&~~ub \\
                       &$12^{+}$&~~10639.6~~&~~10639.0~~&~~ub \\
     $cc\bar{q}\bar{q}$&$10^{+}$&~~ 3725.9~~&~~ 3724.5~~&~~ub \\
                       &$11^{+}$&~~ 3844.3~~&~~ 3843.2~~&~~ub \\
                       &$12^{+}$&~~ 3962.5~~&~~ 3961.0~~&~~ub \\
     $ss\bar{q}\bar{q}$&$10^{+}$&~~  987.0~~&~~  983.5~~&~~ub \\
                       &$11^{+}$&~~ 1408.9~~&~~ 1407.8~~&~~ub \\
                       &$12^{+}$&~~ 1830.5~~&~~ 1827.0~~&~~ub \\
\hline \hline
   \end{tabular}
\end{table}

\subsection{$QQ^{\prime}\bar{q}\bar{q}$ system}

\begin{table}[h]
\centering
\fontsize{9}{9}\selectfont
\caption{The energies of iso-scalar $QQ^{\prime}\bar{q}\bar{q}~(Q,Q^{\prime}=s,c,b)$ systems. ``c.c.(M-M)", ``c.c.(D-A)" and
``c.c.(full)" stand for channel-coupling in meson-meson structure, diquark-antidiquark  structure and two structures mixing (unit: MeV).}\label{QQ3}
\begin{tabular}{ccccccc}
\hline \hline
channel & \multicolumn{2}{c}{$bc\bar{q}\bar{q}$}&\multicolumn{2}{c}{$bs\bar{q}\bar{q}$}&\multicolumn{2}{c}{$cs\bar{q}\bar{q}$}\\ \hline
 \multicolumn{7}{c}{$IJ^P=00^+$} \\ \hline
    $|1111\rangle$ & 7139.8 & $ 41.02\%$ & 5775.5 & $ 93.54\%$ & 2358.1 & $ 95.20\%$ \\
    $|1121\rangle$ & 7518.1 & $  0.10\%$ & 6472.6 & $  0.16\%$ & 3117.9 & $  0.22\%$ \\
    $|1211\rangle$ & 7295.5 & $ 26.92\%$ & 6234.6 & $  9.32\%$ & 2896.1 & $  0.44\%$ \\
    $|1221\rangle$ & 7346.6 & $  0.16\%$ & 6269.4 & $  0.33\%$ & 2871.4 & $  0.01\%$ \\
    $|2133\rangle$ & 7011.5 & $ 31.64\%$ & 6014.6 & $  4.67\%$ & 2660.3 & $  3.20\%$ \\
    $|2243\rangle$ & 7426.1 & $  0.16\%$ & 6391.2 & $  0.94\%$ & 2983.3 & $  0.55\%$ \\
c.c.(M-M)    &\multicolumn{2}{c}{7134.1} &\multicolumn{2}{c}{5774.6}&\multicolumn{2}{c}{2357.2}\\
c.c.(D-A)    &\multicolumn{2}{c}{7005.4} &\multicolumn{2}{c}{5995.3}&\multicolumn{2}{c}{2619.1}\\
c.c.(full)   &\multicolumn{2}{c}{6965.1} &\multicolumn{2}{c}{5756.1}&\multicolumn{2}{c}{2342.1}\\
Threshold    &\multicolumn{2}{c}{7143.5} &\multicolumn{2}{c}{5775.1}&\multicolumn{2}{c}{2356.8}\\\hline
 \multicolumn{7}{c}{$IJ^P=01^+$} \\ \hline
    $|1311\rangle$ & 7259.3 & $ 15.26\%$ & 6195.6 & $  0.16\%$ & 2777.9 & $  0.39\%$ \\
    $|1321\rangle$ & 7517.5 & $  0.01\%$ & 6475.8 & $  0.01\%$ & 3132.8 & $  0.03\%$ \\
    $|1411\rangle$ & 7179.3 & $ 29.95\%$ & 5814.4 & $ 93.11\%$ & 2476.2 & $ 94.49\%$ \\
    $|1421\rangle$ & 7521.2 & $  0.01\%$ & 6478.0 & $  0.15\%$ & 3131.6 & $  0.17\%$ \\
    $|1511\rangle$ & 7298.8 & $ 21.11\%$ & 6234.7 & $  0.25\%$ & 2896.2 & $  0.46\%$ \\
    $|1521\rangle$ & 7431.2 & $  0.01\%$ & 6373.3 & $  0.19\%$ & 3019.8 & $  0.07\%$ \\
    $|2343\rangle$ & 7479.9 & $  0.00\%$ & 6502.7 & $  0.00\%$ & 3134.0 & $  0.19\%$ \\
    $|2433\rangle$ & 7022.3 & $ 33.22\%$ & 6027.9 & $  4.90\%$ & 2694.3 & $  3.75\%$ \\
    $|2543\rangle$ & 7453.1 & $  0.01\%$ & 6448.2 & $  0.71\%$ & 3060.2 & $  0.44\%$ \\
c.c.(M-M)    &\multicolumn{2}{c}{7174.5} &\multicolumn{2}{c}{5813.9}&\multicolumn{2}{c}{2475.9}\\
c.c.(D-A)    &\multicolumn{2}{c}{7020.1} &\multicolumn{2}{c}{6017.9}&\multicolumn{2}{c}{2683.5}\\
c.c.(full)   &\multicolumn{2}{c}{6983.1} &\multicolumn{2}{c}{5797.7}&\multicolumn{2}{c}{2465.3}\\
Threshold    &\multicolumn{2}{c}{7182.3} &\multicolumn{2}{c}{5813.8}&\multicolumn{2}{c}{2474.7}\\\hline
 \multicolumn{7}{c}{$IJ^P=02^+$} \\ \hline
    $|1611\rangle$ & 7297.4 & $ 99.67\%$ & 6234.0 & $ 99.52\%$ & 2895.4 & $ 99.74\%$ \\
    $|1621\rangle$ & 7625.2 & $  0.06\%$ & 6601.7 & $  0.10\%$ & 3262.1 & $  0.20\%$ \\
    $|2643\rangle$ & 7502.8 & $  0.28\%$ & 6545.5 & $  0.38\%$ & 3185.9 & $  0.06\%$ \\
c.c.(M-M)    &\multicolumn{2}{c}{7297.4} &\multicolumn{2}{c}{6233.9}&\multicolumn{2}{c}{2895.4}\\
c.c.(D-A)    &\multicolumn{2}{c}{7502.8} &\multicolumn{2}{c}{6537.6}&\multicolumn{2}{c}{3185.9}\\
c.c.(full)   &\multicolumn{2}{c}{7296.2} &\multicolumn{2}{c}{6233.7}&\multicolumn{2}{c}{2895.2}\\
Threshold    &\multicolumn{2}{c}{7300.1} &\multicolumn{2}{c}{6233.2}&\multicolumn{2}{c}{2894.1}\\\hline
\end{tabular}
\end{table}

\begin{table}[t]
  \centering
   \caption{The energies of iso-vector $QQ^{\prime}\bar{q}\bar{q}~(Q,Q^{\prime}=s,c,b)$ systems.
   'c.c.' stands for full channel coupling (unit: MeV).}\label{QQ4}
  \begin{tabular}{ccccc}
  \hline \hline
    state& $IJ^{P}$& c.c. & Threshold & ~~~status~~~ \\
     $bc\bar{q}\bar{q}$&$10^{+}$&~~ 7144.5~~&~~ 7143.5~~&~~ub \\
                       &$11^{+}$&~~ 7183.5~~&~~ 7182.3~~&~~ub \\
                       &$12^{+}$&~~ 7301.2~~&~~ 7300.1~~&~~ub \\
     $bs\bar{q}\bar{q}$&$10^{+}$&~~ 5777.1~~&~~ 5775.1~~&~~ub \\
                       &$11^{+}$&~~ 5815.9~~&~~ 5813.8~~&~~ub \\
                       &$12^{+}$&~~ 6235.3~~&~~ 6233.2~~&~~ub \\
     $cs\bar{q}\bar{q}$&$10^{+}$&~~ 2359.3~~&~~ 2356.8~~&~~ub \\
                       &$11^{+}$&~~ 2477.2~~&~~ 2474.7~~&~~ub \\
                       &$12^{+}$&~~ 2986.7~~&~~ 2894.1~~&~~ub \\
  \hline \hline
  \end{tabular}
\end{table}

Because $Q$ and $Q^{\prime}$ are different quarks and the $QQ^{\prime}\bar{q}\bar{q}$ system does not have strict symmetry constraint,
all the states with possible quantum numbers are all allowed. The results for iso-scalar and iso-vector $QQ^{\prime}\bar{q}\bar{q}$ states
with all possible quantum numbers are given in Tables \ref{QQ3}-\ref{QQ4}.

For iso-scalar $bc\bar{q}\bar{q}$ systems, the states with quantum number $IJ^P=00^{+}$, $01^{+}$ and $02^{+}$ are all bound
states. In the single channel calculation, the meson-meson states $BD$ with $J=0$, $B^*D$ with $J=1$, $B^*D^*$ with $J=2$ are bound, and
diquark-antidiquark states with ``good" antidiquark and $J=0,1$ are also bound with even larger binding energies. Other states are all
unbound. The channel coupling in one structure lower the energies of the bound states several MeV, generally. For $B^*D^*$ with $J=2$,
the hidden-color channel does not help. The channel coupling with two structures increases the binding energies of the states further.
At last the state with $IJ^P=01^{+}$ has the largest binding energy, $\sim 199$ MeV, which is far smaller than that for $bb\bar{q}\bar{q}$
state with $IJ^P=01^+$, and a little larger than that for $cc\bar{q}\bar{q}$. The phenomenon infers that the attraction, which is provided
by the light antiquark, and are almost the same in these systems, the differences come from the kinetic energy terms, where the different
masses of heavy quark play a role. The state with $IJ^P=00^{+}$ has the second large binding energy, $\sim$ 178 MeV, the states with
$IJ^P=00^{+}$ and $IJ^P=01^{+}$ have the diquark-antidiquark structure, and the state with $IJ^P=02^{+}$ is just bound with binding
energy $\sim 4$ MeV, the spacial structure is approaching the molecular one.
For iso-scalar $bs\bar{q}\bar{q}$ systems, all the energies of states in single channel calculation are above the corresponding thresholds.
For the states with $IJ^P=00^{+}$, the channel coupling in meson-meson structure obtains a bound state with binding energy less than 1 MeV,
the two structures mixing increase the binding energy to 19 MeV. For the states with $IJ^P=01^{+}$, only the channel coupling with two
structures obtains a bound state with binding energy 16 MeV. Two states have the molecular structure. For the states with $IJ^P=02^{+}$,
the channel coupling is not strong enough to form a bound state. The $00^{+}~ bs\bar{q}\bar{q}$ state can be a candidate of $X(5568)$ partner,
which with four different flavor, but with higher mass, $\sim$ 5775 MeV.
For iso-scalar $cs\bar{q}\bar{q}$ systems, we have the similar results with $bs\bar{q}\bar{q}$, only a molecular state with $IJ^P=00^+$,
and a molecular state with $IJ^P=01^+$, are obtained after all channels in two structures are coupled, the binding energies are
$\sim$ 15 MeV and 9 MeV, respectively.

For iso-vector $QQ^{\prime}\bar{q}\bar{q}~(Q,Q^{\prime}=b,c,s)$ systems, all the states are unbound (see Table \ref{QQ4}).

\section{Summary}

In the framework of the chiral constituent quark model, the tetra-quark systems with quark contents $QQ(Q^{\prime})\bar{q}\bar{q}$ are
investigated with the help of GEM. All the possible channels and two kinds of structure are coupled to search for bound states.
The dynamic calculation show that all the iso-vector $QQ(Q^{\prime})\bar{q}\bar{q}$ system are scattering states because
the isospin vector light antiquark are ``bad" ones. In the isospin scalar section, the bound states are abundant.
Due to constraint of symmetry, iso-scalar $QQ\bar{q}\bar{q}$ system with $IJ^P=01^{+}$ is allowed only.
$bb\bar{q}\bar{q}$ and $cc\bar{q}\bar{q}$ are deep bound states with 317.6 MeV and 182.3 MeV binding energy.
$ss\bar{q}\bar{q}$ could be a weak bound state with 9 MeV binding energy. On the other hand, for the iso-scalar
$QQ^{\prime}\bar{q}\bar{q}$ system, the constraint of symmetry is released because of the two different heavy quarks,
more states are allowed. For the $00^{+}$ states, $QQ^{\prime}\bar{q}\bar{q}$ states are all bound states, $bc\bar{q}\bar{q}$ is a deep 
bound state with 178.4 MeV, $bs\bar{q}\bar{q}$ and $cs\bar{q}\bar{q}$ are shallow ones with binding energy 14.7 MeV and 19 MeV,
respectively. $01^{+}$ states are similar to $00^{+}$ states, $bc\bar{q}\bar{q}$ is a deep bound state and $bs\bar{q}\bar{q}$ and
$cs\bar{q}\bar{q}$ are shallow ones. For $02^{+}$ states, only $bc\bar{q}\bar{q}$ is a weak bound state with 3.9 MeV binding energy,
whereas the $bs\bar{q}\bar{q}$ and $cs\bar{q}\bar{q}$ states are all scattering states. Two structures mixing plays an important role
in forming the shallow bound states in $00^{+}$ and $01^{+}$ parts.

From our results, we found that the heavy quark mass has great impact on the binding energy in the $QQ(Q^{'})\bar{q}\bar{q}$ system. 
When the heavy quark mass is more than 1000 MeV, a compact tetraquark can be formed, the $\pi$-exchange potential and color magnetic 
interaction contribute the attraction if the light antidiquark is a ``good" antidiquark. While the heavy quark mass is less than 600 MeV, 
a molecular structure is possible. Is it possible for all light four-quark to form bound states? More calculations are needed.

\end{document}